\documentclass[aps,prb,twocolumn,superscriptaddress]{revtex4}
\usepackage{epsfig}
\begin{document}

\title{\bf Survey of the didemnins: A class of depsipeptide natural products with promising biomedical applications}
\date{February 10, 2005}
\author{Jonathan L. Belof}
\email{jbelof@warp.cas.usf.edu}
\affiliation{Department of Chemistry, University of South Florida\\4202. E. Fowler Ave., CHE205, Tampa, FL 33620-5250}

\begin{abstract}The didemnins represent a versatile class of depsipeptides of marine origin and hold a great deal of potential for biomedical application.  The biological and geographical origins of the didemnins are reviewed in addition to the chemical structures of the major didemnins.  The biological mechanisms behind the antiviral and anticancer effects of selected didemnins are summarized and the special case of dehydrodidemnin B (Aplidin$^{TM}$) is expounded upon including structural characteristics, synthesis, pharmacological mechanism and a discussion of its current clinical trials as an anticancer agent.
\end{abstract}

\maketitle

\section{Introduction}
\label{sec:intro}

The didemnins represent a versatile class of depsipeptides of marine origin and as such they are true natural products.  While their discovery and structural elucidation has been challenging to the researchers involved, the didemnins hold a great deal of potential in pharmacological application.

Since their first isolation from \emph{Trididemnum solidum} 20 years ago, natural product chemists have struggled to determine their structure, origin and biosynthesis.  And while a great deal of knowledge has been gained, the didemnins still remain a mysterious natural product, making this class of cyclic peptides a fascinating subject of chemical research.

The cytotoxic nature of the didemnins was identified immediately upon cell culture screening.  Since their discovery\cite{rinehart1} they have found applications as antibiotic, antiviral and anticancer agents.  The relatively high degree of specificity for their action, combined with newly developed methods of  high-yield synthesis, have made them a perfect  candidate class of compounds to explore for drug discovery. And there are infact several didemnins currently undergoing clinical trials as anticancer agents, the most successful of which thus far has been dehydrodidemnin B (Aplidin$^{TM}$), a subject that is discussed in this review.

\section{Biological and Geographical Origins of the Didemnin Family of Natural Products}
\label{sec:origins}

Many of the didemnins known have been isolated from the Caribbean tunicate \emph{Trididemnum solidum}.  \emph{Trididemnun} belong to the suborder \emph{Aplousobranchia}, order \emph{Enterogona}, class \emph{Ascidiacea}, subphylum \emph{Urochordata}, phylum \emph{Chordata}.  These tunicates (also known as sea squirts) are found in the warm Caribbean waters and have been located in Belize, Columbia, Mexico, Honduras, Panama and Brazil\cite{rinehart1,vervoort1}.  It has been noted that the class \emph{Ascidiacea} contains the only urochordatans that have yielded natural products to date\cite{mcclintock1}.  However, this may merely be due to lack of specimen collection and it is equally possible that natural products will be found in the other classes of \emph{Urochordata}; the fact that \emph{Ascidiacea} is the largest of the classes may also play a role in this over-representation\cite{mcclintock1}.

An interesting case is that of dehydrodidemnin B, which was isolated unexpectedly from the Mediterranean tunicate \emph{Alpidium albicans}\cite{rinehart2}.  The strong relationship between dehydrodidemnin and the other didemnins isolated from the Caribbean tunicates is intriguing given the geographical distance between species.

The ascidiates are typically solitary creatures that remain anchored on the sea floor and may be sparingly motile.  They draw in seawater around them, filter it through mucosal membranes to catch nutrients, and then expel the water back along with any waste products.  Didemnums and Tridemnums are unique in that they may fuse to form clusters of more than one individual organism and may form a single inlet chamber to draw in the surrounding water\cite{margulis1}.

It is unclear whether the origin of the didemnin natural products is ascidian, algal or prokaryotic in nature.  As is frequently the case in marine natural products, determining the exact organism responsible for the secondary metabolites can be exceptionally difficult due to the symbiotic relationships that exist naturally, and also in part due to the ecological interplay between species chemically.  For example, ascidians are known to house prokaryotes such as \emph{Synechocystis} and \emph{Prochloron}, and there are almost certainly other symbionts whose distinct identity has not yet been determined\cite{mcclintock1}.

\section{Structural Characteristics}
\label{sec:structure}

\begin{figure}[htp]
\includegraphics[width=3.0 in]{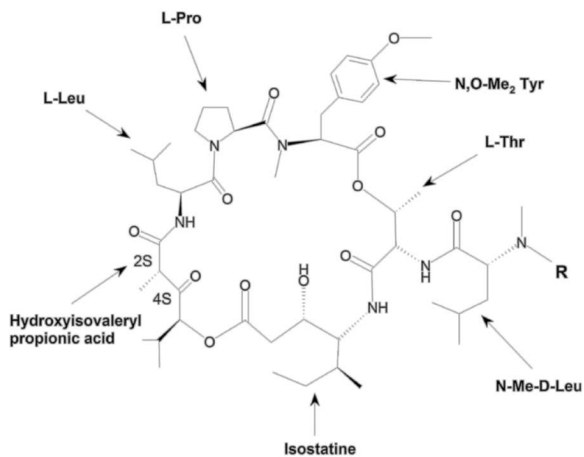}
\caption{The standard Hip/Isostatine macrocycle template for the didemnin family of natural products.  Other variations include norstatine and hydroxyisovaleric acid (Hiv).  The methylated Tyr and Leu residues would appear to indicate enzymatic "tailoring" of the macrocycle.}
\label{fig:figure1}
\end{figure}

The signature characteristic of the didemnins is the depsipeptide macrocycle skeleton.  This depsipeptide cycle is composed of tyrosine-proline-leucine linked to an $\alpha$-($\alpha$-hydroxyisovaleryl)propionyl (Hip) group which in turn is connected to an isostatine unit by an ester bond; on the amide-side of isostatine is a threonine whose hydroxyl side group is ester-linked back to the starting methyl-modified tyrosine, thereby closing the depsipeptide ring.  The N-terminal side of the threonine forms a peptide bond with a methyl-modified leucine; interestingly, this leucine is the D-leucine enantiomer, as opposed to the amino acids of the macrocycle which are all L isomers as typically found in biosynthetic peptides; figure \ref{fig:figure1} depicts the basic macrocycle for the majority of the didemnins (and as explained below, the Hip and isostatine groups may vary depending upon the subtype of didemnin).

Historically, determining the macrocycle structure proved to be challenging.  Initial studies of the didemnin structure were conducted by NMR and chemical methods\cite{rinehart3}, gas chromatography\cite{rinehart4} and the structure and isomerism of the didemnin family was revised several times; one of the greatest impediments is that it was originally believed the isostatine member of the didemnin B macrocycle was actually a statine until X-ray crystallography studies demonstrated otherwise\cite{hossain1}. The stereochemistry of the Hip group also proved to be a challenge to determine, and it was nearly 10 years before the structure of the macrocycle had been confidently settled\cite{banaigs1}.

Past the N-methylated D-leucine lies a variable side-chain that identifies the different didemnins (see figure \ref{fig:table1} below for a summary of the side-chains from some of the didemnins).  Didemnin A, considered to be a structural parent of other derivatives (in terms of it's simplicity and also as a synthetic precursor), has simply a proton for it's side-chain.  The side-chain of didemnin B is a proline-lactic acid residue proper (which in itself is a common side-chain starting point for the majority of the known didemnins); a class of didemnins known as the tamandarins also have this side-chain, however they have a Hiv (hydroxyisovaleric acid) rather than a Hip in the macrocycle\cite{vervoort1}.  In the case of both tamandarin B and nordidemnin B the isostatine has been replaced by a norstatine (norstatine is identical to isostatine except that it lacks a methyl, giving it an isopropyl rather than an isobutyl side-group)\cite{jouin1}.

Didemnins D, E and M are interesting in that their side-chains contain varying lengths of pyroglutamic acid and glutamine following the proline-lactic acid motif\cite{liang1}.  Also described by Rinehart in a patent are didemnins X and Y, which are similar to D, E and M but they end the side-chain with a Hydec (hydroxydecanoic acid), a 10-carbon carboxylic acid.

How could such a unique class of compounds be biosynthesized?  The question is especially intriguing in light of the macrocycle: is the cyclic peptide simply made by ribosomal translation (which would imply a small RNA source template), or could the fragments originate from another cellular source, such as the proteasome, and then the peptide bonds formed by some other enzyme?   If it is formed by mRNA/protein translation, how does one explain the Hip/Hiv and isostatine, or norstatine integral units?  And if the source of the peptide fragments were proteasomal in nature, then would this allow for didemnins to be made in a recombinatorial manner, in response to some event where a foreign protein was degraded?  The presence of the leucine (a D-Leucine isomer no less - highly unusual in living systems) on the side-chain of all the known didemnins would also suggest that the leucine is part of the peptide prior to ring closure by dehydration of the threonine with N,O-Me2-tyrosine.  Does the enzyme responsible simply overcome the steric hinderance involved, or is there another mechanism involving some unknown intermediate?  Unfortunately, very little is known about the biosynthetic pathway from which the didemnins originate, nor the fascinating chemistry that undoubtedly takes place.

\begin{figure}
\begin{tabular}{|l|l|}
\hline
Didemnin A&R $=$ H\\
Didemnin B&R $=$ L-Pro-L-Lac\\
Didemnin C&R $=$ L-Lac\\
Didemnin D&R $=$ L-Pro-L-Lac-(L-Gln)$_3$-L-pGlu\\
Didemnin E&R $=$ L-Pro-L-Lac-(L-Gln)$_2$-L-pGlu\\
Didemnin G&R $=$ CHO\\
Didemnin M&R $=$ L-Pro-L-Lac-L-Gln-L-pGlu\\
Didemnin X&R $=$ L-Pro-L-Lac-(L-Gln)$_3$-Hydec\\
Didemnin Y&R $=$ L-Pro-L-Lac-(L-Gln)$_4$-Hydec\\
Dehydrodidemnin B&R = L-Pro-Pyruvate\\
\hline
\end{tabular}
\caption{R groups that follow the N-Me-D-Leu sidearm for some of the various didemnins that have been discovered to date.  This list is not exhaustive, as there have been more than 40 didemnins discovered and/or synthesized.}
\label{fig:table1}
\end{figure}

\section{Mechanisms of Biological Activity}
\label{sec:mechanism}

Recent studies appear to indicate that the natural function of the didemnins in tunicates is as a fish repellent.  Fluorescent analogs of didemnin B and tamandarin A were tested against the feeding patterns of several species of fish, and were found to alter the predator-prey relationship in favor of the prey\cite{joullie1}.  Why would such a function correspond with biological action against human diseases?  It would appear to be due to the selective cytotoxic potential of the compound: it is not a toxin per se, but rather a functional repellent.  This may give it precisely the balance that is needed for medicinal uses, with the compound being cytotoxic enough to kill vigorously dividing or virally-infected cells, but (ideally) not the host.

One of the earliest biological activities of the didemnins A and B reported were antiviral.  Rinehart noticed the antiviral effects of didemnin on Herpes Simplex Virus 1\cite{rinehart3} and other results soon followed.  It was also determined that \emph{in vitro} antiviral activities against Dengue virus\cite{maldonado1}, Rift Valley fever virus\cite{canonico1}, and Rabies\cite{bussereau1} were enacted in response to didemnin B, with didemnin A being substantially less potent.  It has since been determined that the antiviral effects induced by cell treatment with didemnin A or B are due to inhibition of RNA/DNA/protein synthesis and activation of pro-apoptotic pathways.

Nearly all of the didemnins have demonstrated anticancer effects.  Didemnin A and B were both found to inhibit the growth of L1210 and P388 leukemia tumor cell lines \emph{in vitro}\cite{rinehart3}.  Furthermore, it has been shown that after 2 hours of treatment the halting of cell growth of L1210 cells was irreversible. This was the first clue that something besides protein synthesis was responsible for the cytotoxicity since apoptosis is also irreversible once certain cellular biochemical "checkpoints" have been passed\cite{li1}.  Nordidemnin B was also tested against a rat mammary cancer line and found to alter the signaling pathways of protein kinases involving phosphotidylinositol second messengers, a result that also suggested an unknown process besides (or in conjuction with) protein synthesis inhibition as being responsible for the cytotoxic effects\cite{dominice1}.  It is now know that the didemnins act through at least three different mechanisms: inhibition of protein synthesis, cell-cycle disruption and induction of apoptosis.

Protein synthesis has been shown to be inhibited during peptide elongation.  Didemnin B stabilizes tRNA binding to the ribosome, thereby preventing the next tRNA from entering and elongating the peptide\cite{sirdeshande1}.  This mechanism of cytotoxicity is not new, as several antibiotics (tetracycline, kirromycin) also function in this manner.  Additional work provided by a clever experiment utilizing diphtheria toxin ADP-ribosylation showed that didemnin B forms a complex with EF-1 and the ribosomal subunit\cite{ahuja1}. An early source also shows drastically reduced protein synthesis as a result of the activity of ornithine decarboxylase being terminated by didemnin B\cite{russell1}.

However, in addition to these extensive studies showing that the didemnins inhibit protein synthesis, it was discovered that didemnin B and dehydrodidemnin B both induce apoptosis more rapidly than any other substance reported\cite{beidler1,erba1}.  Follow-up experiments also showed that rapamycin inhibits didemnin B-induced apoptosis, definitively showing that inhibition of protein synthesis is not the sole cause of the apoptosis, but rather that apoptosis is activated by involvement of FK506-binding-protein-25\cite{johnson1}.

\section{Case in Point:  Dehydrodidemnin B (Aplidin$^{TM}$)}
\label{sec:aplidin}

Aplidin$^{TM}$ (dehydrodidemnin B), originally extracted in 1990 from the Mediterranean tunicate \emph{Aplidum albicans} (and since synthesized) is of enormous pharmaceutical value.  This natural product has shown outstanding cytotoxicity against nearly every form of solid-tumor it has been tested against, including non-small cell lung (one of the most deadly and aggressive manifestations of cancer, to which there is currently no effective treatment), prostate sarcoma and melanoma.  Based upon cell culture studies, rudimentary data suggests that the therapeutic index of Aplidin may be high enough to meet onco-drug standards.  Once the effects of cell-cycle arrest and apoptosis were demonstrated \emph{in vitro}, and methods for synthesizing the compound in high yield were determined, Aplidin was put on the fast-track toward becoming a therapeutic drug and is currently in Phase II clinical trials.

\subsection{Chemical Structure}
\label{sec:aplidin_structure}

\begin{figure}[htp]
\includegraphics[width=3.0 in]{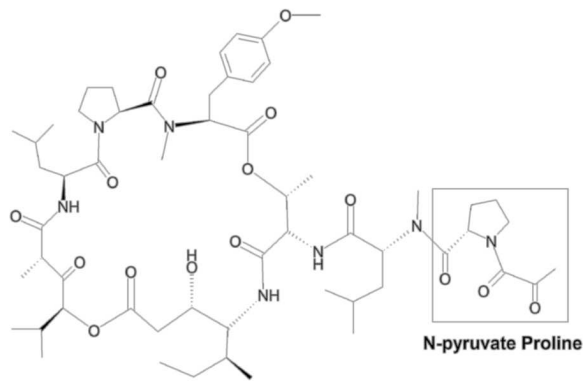}
\caption{Structural diagram of dehydrodidemnin B.  The standard macrocycle is in place (Hip, isostatine) with a pyruvylproline side-chain.  Interestingly, this side-chain is only slightly different than the lactate side-chain of didemnin B, yet the biological effects differ between the two depsipeptides significantly.}
\label{fig:figure2}
\end{figure}

Dehydrodidemnin B is structurally similar to didemnin B; instead of an N-linked lactate there is a pyruvate on the proline side-chain.  The macrocycle peptide is the same as described earlier in this review (Hip, isostatine) and appears essential to maintaining it's biological activity.

It is not known why, biochemically speaking, the pyruvyl-proline side-chain would make Aplidin a better candidate over didemnin B in terms of both cytotoxicity against cancer cells and also in lacking the cardiovascular side-effects which caused didemnin B to prematurely terminate it's Phase I trial.

\subsection{Laboratory Synthesis}
\label{sec:aplidin_synthesis}

Synthesis of Dehydrodidemnin B is a simple, single step away from didemnin A involving attachment of the pyruvylproline to the N-Me-D-Leu site.  Since this essentially boils down to the synthesis of didemnin A, the focus of the synthesis is overcoming the difficulties in forming the macrocycle.

\begin{figure}[htp]
\includegraphics[width=3.0 in]{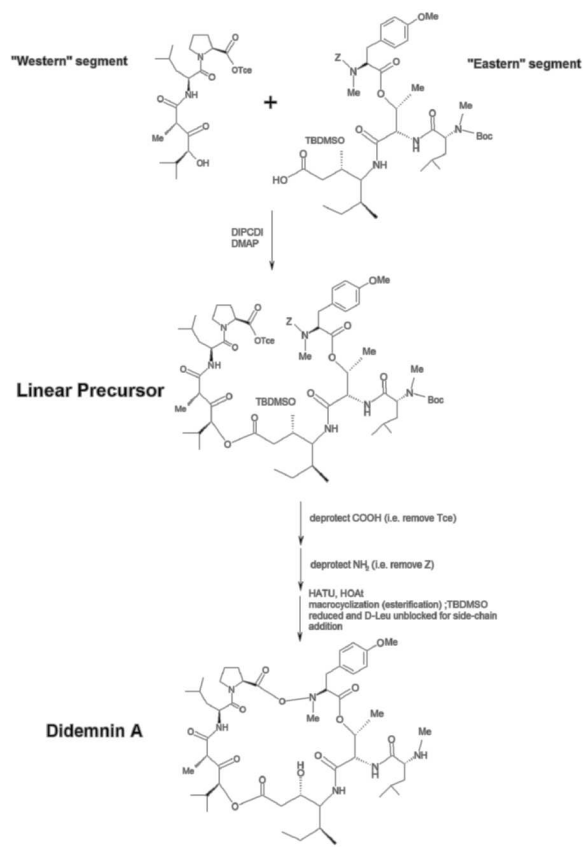}
\caption{Esterfication of the "eastern" and "western" segments into a linear macrocycle precursor by oxidation with DCC or DIPCDI in the presence of DMAP in refluxing chloroform.  Deprotection of the carboxyl group by addition of Zinc dust in THF followed by NH$_4$OAc.  Sequential deprotection of the amino group by 10 \% Pd-C in THF.  Finally, macrocyclization in HATU, HOAt and DIEA.  Each step involves additional washes, filtrations and either silica gel or HPLC for further purification of product.}
\label{fig:figure3}
\end{figure}

Several synthetic routes for the didemnins have been published, but here we will explore the formation of the macrocycle as ring closure of a linear precursor by way of macrolactamization\cite{jou1}, the following synthetic scheme being the method of Jou et al.  There are several possible points at which the ring can be closed, but in this route the N,O-Me2-Tyr to L-Pro was selected since there are drawbacks to the various other options (for instance, Hip to Leu would have the tendency of decarboxylating the -keto acid formed).  Macrocyclization through this selected route should not yield any obvious unwanted side reactions.

The major difficulties to be overcome are (a) the lack of a good nucleophile in forming the ester linkage without affecting the carboxylic acid and (b) overcoming steric hinderance from the crowding of peptide bonds around a constrained ring.  This synthetic route makes use of phosphonium and uronium salt-based reagents that have been developed for such purposes of constrained amino acid chemistry.

The synthesis begins with the formation of "eastern" and "western" segments; both segments are formed in conjuction with the use of Boc-, Z- as amino group protectors, Bzl and TBDMS ethers for alcohol protectors, and SEM esters for the protection of the carboxylic acids (synthesis of these precursors can be found in the literature\cite{jou1,li2}).  Both segments are then linked by an ester bond that is formed through hydrolysis to give the linear precursor from which cyclization will give didemnin A.  Esterfication takes place by use of DIPCDI and DMAP in refluxing chloroform with careful control of temperature.  This final step will give the linear macrocycle precursor in 60 \% yield after column chromatography.

\begin{figure}[htp]
\includegraphics[width=3.0 in]{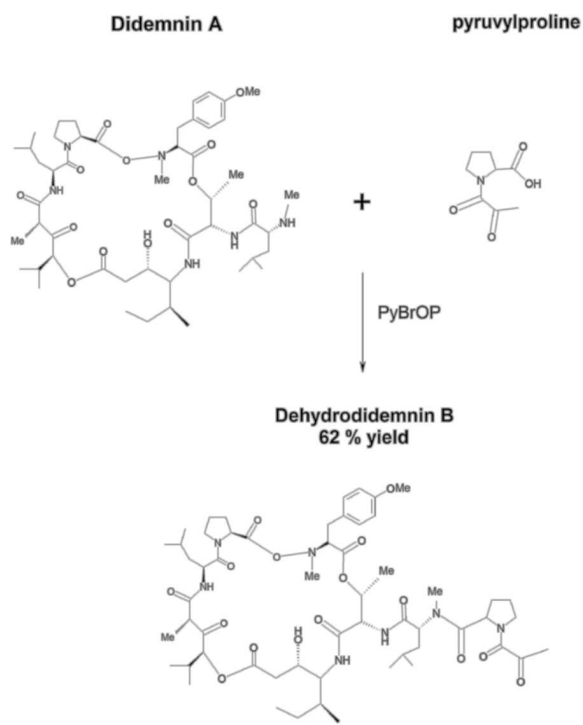}
\caption{Final reaction to give dehydrodidemnin B.  Dichloromethane, PyBrOP and DIEA added, product extracted in AcOEt.  Washed in salt and potassium sulfate solutions, filtered, dried and purified by silica gel chromatography.}
\label{fig:figure4}
\end{figure}

Conversion of the linear precursor into didemnin A takes place through deprotection of the carboxyl terminus, yielding a first intermediate followed by deprotection of the amino terminus and then finally initiating macrocyclization by use of HATU in HOAt.  Quenching with 5 M HCl-dioxane gives didemnin A in approximately 28 \% yield for this macrocyclization step (see figure \ref{fig:figure3}).

Final coupling of pyruvylproline to the side-chain D-Leu of didemnin A is done through the use of bromo-tris-pyrrolidino-phosphonium hexafluorophosphate (PyBrOP), an excellent reagent for coupling N-Me amino acids with enantiomeric specificity, giving the final product dehydrodidemnin B in 62 \% yield (see figure \ref{fig:figure4}).

\subsection{Biological Mechanism of Action, Biomedical Applications and Clinical Trials}
\label{sec:aplidin_mechanism}

Dehydrodidemnin B causes cell-cycle arrest at the G1-S checkpoint and also at G2-M by a unique mechanism\cite{erba1}.  Inhibition of JNK and p38 MAP kinase blocked the actions of Aplidin, showing that the apoptotic program was being triggered somewhere upstream of the caspases.  This mechanism most likely involves PKC delta, an important protein kinase that has in recent years been shown to be an active player in apoptosis\cite{garcia1}.

It has very recently been determined that a direct target of Aplidin is EGFR (epidermal growth factor receptor).  Ironically, EGFR has been identified as a factor in promoting tumor growth.  It appears that dehydrodidemnin B binds and activates EGFR continuously which induces the apoptotic program\cite{cuadrado1}.  Along similar lines, Aplidin inhibits the secretion of VEGF (vascular epidermal growth factor) which has been shown to kill the leukemia cell line MOLT-4 and presumably could have applications in suppressing angiogenesis\cite{broggini1}.

Given the initial cell culture screening and the biological mechanism of action supported by further biochemical testing, dehydrodidemnin B has been identified as a prime candidate for biomedical application in treating cancer and autoimmune diseases.  PharmaMar S.A. (Madrid, Spain) is currently conducting Phase II clinical trials of Aplidin against solid tumors and hematological malignancies (Multiple Myeloma, Hodgkin Lymphoma, Acute Lymphoblastic Leukemia), with approximately 400 patients having already been treated in the USA, Canada and Europe.

\section{Conclusions}
\label{sec:conclusions}

The didemnins are a fascinating family of natural products that have a bright future.  They form a conceptual bridge between the traditional small-molecule drugs and the more modern drugs derived from peptides and monoclonal antibodies. They have demonstrated the potent action and versatility of a small-molecule and yet, due to the macrocycle component, can interact more specifically with proteins that comprise an integral part of cellular processes.

It is hoped that the biosynthetic pathways that give rise to the didemnins will be elucidated in the future.  Such a discovery would not only enhance our scientific knowledge of biosynthetic natural product chemistry, but could also open the door to creating recombinant biosynthetic systems (such as bioengineered bacteria) that could be used in high throughput screening and research of depsipeptide drugs.

\bibliography{didemnin_review}

\begin{thebibliography}{28}
\expandafter\ifx\csname natexlab\endcsname\relax\def\natexlab#1{#1}\fi
\expandafter\ifx\csname bibnamefont\endcsname\relax
  \def\bibnamefont#1{#1}\fi
\expandafter\ifx\csname bibfnamefont\endcsname\relax
  \def\bibfnamefont#1{#1}\fi
\expandafter\ifx\csname citenamefont\endcsname\relax
  \def\citenamefont#1{#1}\fi
\expandafter\ifx\csname url\endcsname\relax
  \def\url#1{\texttt{#1}}\fi
\expandafter\ifx\csname urlprefix\endcsname\relax\def\urlprefix{URL }\fi
\providecommand{\bibinfo}[2]{#2}
\providecommand{\eprint}[2][]{\url{#2}}

\bibitem[{\citenamefont{Rinehart{, Jr.}}(1985)}]{rinehart1}
\bibinfo{author}{\bibfnamefont{K.}~\bibnamefont{Rinehart{, Jr.}}},
  \emph{\bibinfo{title}{{U.S. Patent No. 4,493,796}}} (\bibinfo{year}{1985}).

\bibitem[{\citenamefont{Vervoort et~al.}(2000)\citenamefont{Vervoort, Fenical,
  and Epifanio}}]{vervoort1}
\bibinfo{author}{\bibfnamefont{H.}~\bibnamefont{Vervoort}},
  \bibinfo{author}{\bibfnamefont{W.}~\bibnamefont{Fenical}}, \bibnamefont{and}
  \bibinfo{author}{\bibfnamefont{R.}~\bibnamefont{Epifanio}},
  \bibinfo{journal}{J.Org.Chem.} \textbf{\bibinfo{volume}{65}},
  \bibinfo{pages}{782} (\bibinfo{year}{2000}).

\bibitem[{\citenamefont{Mc{C}lintock and Baker}(2001)}]{mcclintock1}
\bibinfo{author}{\bibfnamefont{J.}~\bibnamefont{Mc{C}lintock}}
  \bibnamefont{and} \bibinfo{author}{\bibfnamefont{B.}~\bibnamefont{Baker}},
  \emph{\bibinfo{title}{Marine Chemical Ecology}} (\bibinfo{publisher}{CRC
  Press}, \bibinfo{address}{Boca Raton, FL}, \bibinfo{year}{2001}).

\bibitem[{\citenamefont{Rinehart{, Jr.} and
  Lithgow-Bertelloni}(1998)}]{rinehart2}
\bibinfo{author}{\bibfnamefont{K.}~\bibnamefont{Rinehart{, Jr.}}}
  \bibnamefont{and}
  \bibinfo{author}{\bibfnamefont{A.}~\bibnamefont{Lithgow-Bertelloni}},
  \emph{\bibinfo{title}{{U.S. Patent No. 5,834,586}}} (\bibinfo{year}{1998}).

\bibitem[{\citenamefont{Margulis and Schwartz}(1982)}]{margulis1}
\bibinfo{author}{\bibfnamefont{L.}~\bibnamefont{Margulis}} \bibnamefont{and}
  \bibinfo{author}{\bibfnamefont{K.}~\bibnamefont{Schwartz}},
  \emph{\bibinfo{title}{Five Kingdoms}} (\bibinfo{publisher}{W.H. Freeman},
  \bibinfo{address}{San Francisco}, \bibinfo{year}{1982}).

\bibitem[{\citenamefont{Rinehart{, Jr.} et~al.}(1998)\citenamefont{Rinehart{,
  Jr.}, Gloer, Cook{, Jr.}, Mizsak, and Scahill}}]{rinehart3}
\bibinfo{author}{\bibfnamefont{K.}~\bibnamefont{Rinehart{, Jr.}}},
  \bibinfo{author}{\bibfnamefont{J.}~\bibnamefont{Gloer}},
  \bibinfo{author}{\bibfnamefont{J.}~\bibnamefont{Cook{, Jr.}}},
  \bibinfo{author}{\bibfnamefont{S.}~\bibnamefont{Mizsak}}, \bibnamefont{and}
  \bibinfo{author}{\bibfnamefont{T.}~\bibnamefont{Scahill}},
  \bibinfo{journal}{J. Am. Chem. Soc.} \textbf{\bibinfo{volume}{103}},
  \bibinfo{pages}{1857} (\bibinfo{year}{1998}).

\bibitem[{\citenamefont{Rinehart{, Jr.} et~al.}(1983)\citenamefont{Rinehart{,
  Jr.}, Gloer, Wilson, Hughes{, Jr.}, Li, Renis, and Mc{G}ovren}}]{rinehart4}
\bibinfo{author}{\bibfnamefont{K.}~\bibnamefont{Rinehart{, Jr.}}},
  \bibinfo{author}{\bibfnamefont{J.}~\bibnamefont{Gloer}},
  \bibinfo{author}{\bibfnamefont{G.}~\bibnamefont{Wilson}},
  \bibinfo{author}{\bibfnamefont{R.}~\bibnamefont{Hughes{, Jr.}}},
  \bibinfo{author}{\bibfnamefont{L.}~\bibnamefont{Li}},
  \bibinfo{author}{\bibfnamefont{H.}~\bibnamefont{Renis}}, \bibnamefont{and}
  \bibinfo{author}{\bibfnamefont{J.}~\bibnamefont{Mc{G}ovren}},
  \bibinfo{journal}{Fed. Proc.} \textbf{\bibinfo{volume}{42}},
  \bibinfo{pages}{87} (\bibinfo{year}{1983}).

\bibitem[{\citenamefont{Hossain et~al.}(1988)\citenamefont{Hossain, der Helm,
  Antel, Sheldrick, Sanduja, and Weinheimer}}]{hossain1}
\bibinfo{author}{\bibfnamefont{M.}~\bibnamefont{Hossain}},
  \bibinfo{author}{\bibfnamefont{D.~V.} \bibnamefont{der Helm}},
  \bibinfo{author}{\bibfnamefont{J.}~\bibnamefont{Antel}},
  \bibinfo{author}{\bibfnamefont{G.}~\bibnamefont{Sheldrick}},
  \bibinfo{author}{\bibfnamefont{S.}~\bibnamefont{Sanduja}}, \bibnamefont{and}
  \bibinfo{author}{\bibfnamefont{A.}~\bibnamefont{Weinheimer}},
  \bibinfo{journal}{Proc. Natl. Acad. Sci.} \textbf{\bibinfo{volume}{85}},
  \bibinfo{pages}{4118} (\bibinfo{year}{1988}).

\bibitem[{\citenamefont{Banaigs et~al.}(1989)\citenamefont{Banaigs, Jeanty,
  Francisco, Jouin, Poncet, Heitz, Cave, Prome, Wahl, and Lafargue}}]{banaigs1}
\bibinfo{author}{\bibfnamefont{B.}~\bibnamefont{Banaigs}},
  \bibinfo{author}{\bibfnamefont{G.}~\bibnamefont{Jeanty}},
  \bibinfo{author}{\bibfnamefont{C.}~\bibnamefont{Francisco}},
  \bibinfo{author}{\bibfnamefont{P.}~\bibnamefont{Jouin}},
  \bibinfo{author}{\bibfnamefont{J.}~\bibnamefont{Poncet}},
  \bibinfo{author}{\bibfnamefont{A.}~\bibnamefont{Heitz}},
  \bibinfo{author}{\bibfnamefont{A.}~\bibnamefont{Cave}},
  \bibinfo{author}{\bibfnamefont{J.}~\bibnamefont{Prome}},
  \bibinfo{author}{\bibfnamefont{M.}~\bibnamefont{Wahl}}, \bibnamefont{and}
  \bibinfo{author}{\bibfnamefont{F.}~\bibnamefont{Lafargue}},
  \bibinfo{journal}{Tetrahedron} \textbf{\bibinfo{volume}{45}},
  \bibinfo{pages}{181} (\bibinfo{year}{1989}).

\bibitem[{\citenamefont{Jouin et~al.}(1991)\citenamefont{Jouin, Poncet, Dufour,
  Aumelas, and Pantaloni}}]{jouin1}
\bibinfo{author}{\bibfnamefont{P.}~\bibnamefont{Jouin}},
  \bibinfo{author}{\bibfnamefont{J.}~\bibnamefont{Poncet}},
  \bibinfo{author}{\bibfnamefont{M.}~\bibnamefont{Dufour}},
  \bibinfo{author}{\bibfnamefont{A.}~\bibnamefont{Aumelas}}, \bibnamefont{and}
  \bibinfo{author}{\bibfnamefont{A.}~\bibnamefont{Pantaloni}},
  \bibinfo{journal}{J. Med. Chem.} \textbf{\bibinfo{volume}{34}},
  \bibinfo{pages}{486} (\bibinfo{year}{1991}).

\bibitem[{\citenamefont{Liang et~al.}(2000)\citenamefont{Liang, Vera, and
  Joullie}}]{liang1}
\bibinfo{author}{\bibfnamefont{B.}~\bibnamefont{Liang}},
  \bibinfo{author}{\bibfnamefont{M.}~\bibnamefont{Vera}}, \bibnamefont{and}
  \bibinfo{author}{\bibfnamefont{M.}~\bibnamefont{Joullie}},
  \bibinfo{journal}{J. Org. Chem.} \textbf{\bibinfo{volume}{65}},
  \bibinfo{pages}{4762} (\bibinfo{year}{2000}).

\bibitem[{\citenamefont{Joullie et~al.}(2003)\citenamefont{Joullie, Leonard,
  Portonovo, Liang, Ding, and {La Clair}}}]{joullie1}
\bibinfo{author}{\bibfnamefont{M.}~\bibnamefont{Joullie}},
  \bibinfo{author}{\bibfnamefont{M.}~\bibnamefont{Leonard}},
  \bibinfo{author}{\bibfnamefont{P.}~\bibnamefont{Portonovo}},
  \bibinfo{author}{\bibfnamefont{B.}~\bibnamefont{Liang}},
  \bibinfo{author}{\bibfnamefont{X.}~\bibnamefont{Ding}}, \bibnamefont{and}
  \bibinfo{author}{\bibfnamefont{J.}~\bibnamefont{{La Clair}}},
  \bibinfo{journal}{Bioconjug. Chem.} \textbf{\bibinfo{volume}{14}},
  \bibinfo{pages}{30} (\bibinfo{year}{2003}).

\bibitem[{\citenamefont{Maldonado et~al.}(1982)\citenamefont{Maldonado,
  Lavergne, and Kraiselburd}}]{maldonado1}
\bibinfo{author}{\bibfnamefont{E.}~\bibnamefont{Maldonado}},
  \bibinfo{author}{\bibfnamefont{J.}~\bibnamefont{Lavergne}}, \bibnamefont{and}
  \bibinfo{author}{\bibfnamefont{E.}~\bibnamefont{Kraiselburd}},
  \bibinfo{journal}{P.R. Health Sci.} \textbf{\bibinfo{volume}{1}},
  \bibinfo{pages}{22} (\bibinfo{year}{1982}).

\bibitem[{\citenamefont{Canonico et~al.}(1982)\citenamefont{Canonico, Pannier,
  Huggins, and Rinehart}}]{canonico1}
\bibinfo{author}{\bibfnamefont{P.}~\bibnamefont{Canonico}},
  \bibinfo{author}{\bibfnamefont{W.}~\bibnamefont{Pannier}},
  \bibinfo{author}{\bibfnamefont{K.}~\bibnamefont{Huggins}}, \bibnamefont{and}
  \bibinfo{author}{\bibfnamefont{K.}~\bibnamefont{Rinehart}},
  \bibinfo{journal}{Antimicrob. Agents Chemother.}
  \textbf{\bibinfo{volume}{22}}, \bibinfo{pages}{696} (\bibinfo{year}{1982}).

\bibitem[{\citenamefont{Bussereau et~al.}(1988)\citenamefont{Bussereau, Picard,
  Blancou, and Sureau}}]{bussereau1}
\bibinfo{author}{\bibfnamefont{F.}~\bibnamefont{Bussereau}},
  \bibinfo{author}{\bibfnamefont{M.}~\bibnamefont{Picard}},
  \bibinfo{author}{\bibfnamefont{J.}~\bibnamefont{Blancou}}, \bibnamefont{and}
  \bibinfo{author}{\bibfnamefont{P.}~\bibnamefont{Sureau}},
  \bibinfo{journal}{Acta Virol.} \textbf{\bibinfo{volume}{32}},
  \bibinfo{pages}{33} (\bibinfo{year}{1988}).

\bibitem[{\citenamefont{L.H.~Li et~al.}(1984)\citenamefont{L.H.~Li, Wallace,
  Krueger, Prairie, and Im}}]{li1}
\bibinfo{author}{\bibfnamefont{L.~T.} \bibnamefont{L.H.~Li}},
  \bibinfo{author}{\bibfnamefont{T.}~\bibnamefont{Wallace}},
  \bibinfo{author}{\bibfnamefont{W.}~\bibnamefont{Krueger}},
  \bibinfo{author}{\bibfnamefont{M.}~\bibnamefont{Prairie}}, \bibnamefont{and}
  \bibinfo{author}{\bibfnamefont{W.}~\bibnamefont{Im}},
  \bibinfo{journal}{Cancer Lett.} \textbf{\bibinfo{volume}{23}},
  \bibinfo{pages}{279} (\bibinfo{year}{1984}).

\bibitem[{\citenamefont{C.~Dominice et~al.}(1994)\citenamefont{C.~Dominice,
  Dufour, Patino, Maanzoni, Grazzini, Jonin, and Guillon}}]{dominice1}
\bibinfo{author}{\bibfnamefont{C.}~\bibnamefont{C.~Dominice}},
  \bibinfo{author}{\bibfnamefont{M.}~\bibnamefont{Dufour}},
  \bibinfo{author}{\bibfnamefont{N.}~\bibnamefont{Patino}},
  \bibinfo{author}{\bibfnamefont{O.}~\bibnamefont{Maanzoni}},
  \bibinfo{author}{\bibfnamefont{E.}~\bibnamefont{Grazzini}},
  \bibinfo{author}{\bibfnamefont{P.}~\bibnamefont{Jonin}}, \bibnamefont{and}
  \bibinfo{author}{\bibfnamefont{G.}~\bibnamefont{Guillon}},
  \bibinfo{journal}{J. Pharm. Exp. Ther.} \textbf{\bibinfo{volume}{271}},
  \bibinfo{pages}{107} (\bibinfo{year}{1994}).

\bibitem[{\citenamefont{SirDeshpande and Toogood}(1995)}]{sirdeshande1}
\bibinfo{author}{\bibfnamefont{B.}~\bibnamefont{SirDeshpande}}
  \bibnamefont{and} \bibinfo{author}{\bibfnamefont{P.}~\bibnamefont{Toogood}},
  \bibinfo{journal}{Biochemistry} \textbf{\bibinfo{volume}{34}},
  \bibinfo{pages}{9177} (\bibinfo{year}{1995}).

\bibitem[{\citenamefont{Ahuja et~al.}(2000)\citenamefont{Ahuja, Vera,
  SirDeshpande, Morimoto, Williams, Joullie, and Toogood}}]{ahuja1}
\bibinfo{author}{\bibfnamefont{D.}~\bibnamefont{Ahuja}},
  \bibinfo{author}{\bibfnamefont{M.}~\bibnamefont{Vera}},
  \bibinfo{author}{\bibfnamefont{B.}~\bibnamefont{SirDeshpande}},
  \bibinfo{author}{\bibfnamefont{H.}~\bibnamefont{Morimoto}},
  \bibinfo{author}{\bibfnamefont{P.}~\bibnamefont{Williams}},
  \bibinfo{author}{\bibfnamefont{M.}~\bibnamefont{Joullie}}, \bibnamefont{and}
  \bibinfo{author}{\bibfnamefont{P.}~\bibnamefont{Toogood}},
  \bibinfo{journal}{Biochemistry} \textbf{\bibinfo{volume}{39}},
  \bibinfo{pages}{4339} (\bibinfo{year}{2000}).

\bibitem[{\citenamefont{Russell et~al.}(1987)\citenamefont{Russell, Buckley,
  Montgomery, Larson, Gout, Beer, Putnam, Zukoski, and Kibler}}]{russell1}
\bibinfo{author}{\bibfnamefont{D.}~\bibnamefont{Russell}},
  \bibinfo{author}{\bibfnamefont{A.}~\bibnamefont{Buckley}},
  \bibinfo{author}{\bibfnamefont{D.}~\bibnamefont{Montgomery}},
  \bibinfo{author}{\bibfnamefont{N.}~\bibnamefont{Larson}},
  \bibinfo{author}{\bibfnamefont{P.}~\bibnamefont{Gout}},
  \bibinfo{author}{\bibfnamefont{C.}~\bibnamefont{Beer}},
  \bibinfo{author}{\bibfnamefont{C.}~\bibnamefont{Putnam}},
  \bibinfo{author}{\bibfnamefont{C.}~\bibnamefont{Zukoski}}, \bibnamefont{and}
  \bibinfo{author}{\bibfnamefont{R.}~\bibnamefont{Kibler}},
  \bibinfo{journal}{J. Immunol.} \textbf{\bibinfo{volume}{138}},
  \bibinfo{pages}{276} (\bibinfo{year}{1987}).

\bibitem[{\citenamefont{Beidler et~al.}(1999)\citenamefont{Beidler, Ahuja,
  Wicha, and Toogood}}]{beidler1}
\bibinfo{author}{\bibfnamefont{D.}~\bibnamefont{Beidler}},
  \bibinfo{author}{\bibfnamefont{D.}~\bibnamefont{Ahuja}},
  \bibinfo{author}{\bibfnamefont{M.}~\bibnamefont{Wicha}}, \bibnamefont{and}
  \bibinfo{author}{\bibfnamefont{P.}~\bibnamefont{Toogood}},
  \bibinfo{journal}{Biochem. Pharmacol.} \textbf{\bibinfo{volume}{58}},
  \bibinfo{pages}{1067} (\bibinfo{year}{1999}).

\bibitem[{\citenamefont{Erba et~al.}(2002)\citenamefont{Erba, Bassano,
  DiLiberti, Muradore, Chiorino, Ubezio, Vignati, Codegoni, Desiderio,
  Faircloth et~al.}}]{erba1}
\bibinfo{author}{\bibfnamefont{E.}~\bibnamefont{Erba}},
  \bibinfo{author}{\bibfnamefont{L.}~\bibnamefont{Bassano}},
  \bibinfo{author}{\bibfnamefont{G.}~\bibnamefont{DiLiberti}},
  \bibinfo{author}{\bibfnamefont{I.}~\bibnamefont{Muradore}},
  \bibinfo{author}{\bibfnamefont{G.}~\bibnamefont{Chiorino}},
  \bibinfo{author}{\bibfnamefont{P.}~\bibnamefont{Ubezio}},
  \bibinfo{author}{\bibfnamefont{S.}~\bibnamefont{Vignati}},
  \bibinfo{author}{\bibfnamefont{A.}~\bibnamefont{Codegoni}},
  \bibinfo{author}{\bibfnamefont{M.}~\bibnamefont{Desiderio}},
  \bibinfo{author}{\bibfnamefont{G.}~\bibnamefont{Faircloth}},
  \bibnamefont{et~al.}, \bibinfo{journal}{Br. J. Cancer.}
  \textbf{\bibinfo{volume}{86}}, \bibinfo{pages}{1510} (\bibinfo{year}{2002}).

\bibitem[{\citenamefont{Johnson and Lawen}(1999)}]{johnson1}
\bibinfo{author}{\bibfnamefont{K.}~\bibnamefont{Johnson}} \bibnamefont{and}
  \bibinfo{author}{\bibfnamefont{A.}~\bibnamefont{Lawen}},
  \bibinfo{journal}{Immunol. Cell Biol.} \textbf{\bibinfo{volume}{7}},
  \bibinfo{pages}{242} (\bibinfo{year}{1999}).

\bibitem[{\citenamefont{Jou et~al.}(1997)\citenamefont{Jou, Gonzalez,
  Albericio, Lloyd-Williams, and Giralt}}]{jou1}
\bibinfo{author}{\bibfnamefont{G.}~\bibnamefont{Jou}},
  \bibinfo{author}{\bibfnamefont{I.}~\bibnamefont{Gonzalez}},
  \bibinfo{author}{\bibfnamefont{F.}~\bibnamefont{Albericio}},
  \bibinfo{author}{\bibfnamefont{P.}~\bibnamefont{Lloyd-Williams}},
  \bibnamefont{and} \bibinfo{author}{\bibfnamefont{E.}~\bibnamefont{Giralt}},
  \bibinfo{journal}{J. Org. Chem.} \textbf{\bibinfo{volume}{62}},
  \bibinfo{pages}{354} (\bibinfo{year}{1997}).

\bibitem[{\citenamefont{Li et~al.}(1990)\citenamefont{Li, Ewing, Harris, and
  Joullie}}]{li2}
\bibinfo{author}{\bibfnamefont{W.}~\bibnamefont{Li}},
  \bibinfo{author}{\bibfnamefont{W.}~\bibnamefont{Ewing}},
  \bibinfo{author}{\bibfnamefont{B.}~\bibnamefont{Harris}}, \bibnamefont{and}
  \bibinfo{author}{\bibfnamefont{M.}~\bibnamefont{Joullie}},
  \bibinfo{journal}{J. Am. Chem. Soc.} \textbf{\bibinfo{volume}{112}},
  \bibinfo{pages}{7659} (\bibinfo{year}{1990}).

\bibitem[{\citenamefont{Garcia-Fernandez
  et~al.}(2002)\citenamefont{Garcia-Fernandez, Losada, Alcaide, Alvarez,
  Cuadrado, Gonzalez, Nakayama, Nakayama, Fernandez-Sousa, Munoz
  et~al.}}]{garcia1}
\bibinfo{author}{\bibfnamefont{L.}~\bibnamefont{Garcia-Fernandez}},
  \bibinfo{author}{\bibfnamefont{A.}~\bibnamefont{Losada}},
  \bibinfo{author}{\bibfnamefont{V.}~\bibnamefont{Alcaide}},
  \bibinfo{author}{\bibfnamefont{A.}~\bibnamefont{Alvarez}},
  \bibinfo{author}{\bibfnamefont{A.}~\bibnamefont{Cuadrado}},
  \bibinfo{author}{\bibfnamefont{L.}~\bibnamefont{Gonzalez}},
  \bibinfo{author}{\bibfnamefont{K.}~\bibnamefont{Nakayama}},
  \bibinfo{author}{\bibfnamefont{K.}~\bibnamefont{Nakayama}},
  \bibinfo{author}{\bibfnamefont{J.}~\bibnamefont{Fernandez-Sousa}},
  \bibinfo{author}{\bibfnamefont{A.}~\bibnamefont{Munoz}},
  \bibnamefont{et~al.}, \bibinfo{journal}{Oncogene}
  \textbf{\bibinfo{volume}{21}}, \bibinfo{pages}{7533} (\bibinfo{year}{2002}).

\bibitem[{\citenamefont{Cuadrado et~al.}(2003)\citenamefont{Cuadrado,
  Garcia-Fernandez, Gonzalez, Suarez, Losada, Alcaide, Martinez,
  Fernandez-Sousa, Sanchez-Puelles, and Munoz}}]{cuadrado1}
\bibinfo{author}{\bibfnamefont{A.}~\bibnamefont{Cuadrado}},
  \bibinfo{author}{\bibfnamefont{L.}~\bibnamefont{Garcia-Fernandez}},
  \bibinfo{author}{\bibfnamefont{L.}~\bibnamefont{Gonzalez}},
  \bibinfo{author}{\bibfnamefont{Y.}~\bibnamefont{Suarez}},
  \bibinfo{author}{\bibfnamefont{A.}~\bibnamefont{Losada}},
  \bibinfo{author}{\bibfnamefont{V.}~\bibnamefont{Alcaide}},
  \bibinfo{author}{\bibfnamefont{T.}~\bibnamefont{Martinez}},
  \bibinfo{author}{\bibfnamefont{J.}~\bibnamefont{Fernandez-Sousa}},
  \bibinfo{author}{\bibfnamefont{J.}~\bibnamefont{Sanchez-Puelles}},
  \bibnamefont{and} \bibinfo{author}{\bibfnamefont{A.}~\bibnamefont{Munoz}},
  \bibinfo{journal}{J. Biol. Chem.} \textbf{\bibinfo{volume}{278}},
  \bibinfo{pages}{241} (\bibinfo{year}{2003}).

\bibitem[{\citenamefont{Broggini et~al.}(2003)\citenamefont{Broggini, Marchini,
  Galliera, Borsotti, Taraboletti, Erba, Sironi, Jimeno, Faircloth, Giavazzi
  et~al.}}]{broggini1}
\bibinfo{author}{\bibfnamefont{M.}~\bibnamefont{Broggini}},
  \bibinfo{author}{\bibfnamefont{S.}~\bibnamefont{Marchini}},
  \bibinfo{author}{\bibfnamefont{E.}~\bibnamefont{Galliera}},
  \bibinfo{author}{\bibfnamefont{P.}~\bibnamefont{Borsotti}},
  \bibinfo{author}{\bibfnamefont{G.}~\bibnamefont{Taraboletti}},
  \bibinfo{author}{\bibfnamefont{E.}~\bibnamefont{Erba}},
  \bibinfo{author}{\bibfnamefont{M.}~\bibnamefont{Sironi}},
  \bibinfo{author}{\bibfnamefont{J.}~\bibnamefont{Jimeno}},
  \bibinfo{author}{\bibfnamefont{G.}~\bibnamefont{Faircloth}},
  \bibinfo{author}{\bibfnamefont{R.}~\bibnamefont{Giavazzi}},
  \bibnamefont{et~al.}, \bibinfo{journal}{Leukemia}
  \textbf{\bibinfo{volume}{17}}, \bibinfo{pages}{52} (\bibinfo{year}{2003}).

\end{thebibliography}

\end{document}